\documentclass[12pt]{iopart}
\usepackage{graphicx}
\usepackage{cite}
\usepackage[justification=centering]{caption}
\begin{document}

\title[Hidden symmetries and nonlinear constitutive relations]{Hidden symmetries and nonlinear constitutive relations for tight-coupling heat engines}

\author{S Q Sheng and Z C Tu\footnote{Author to whom any correspondence should be addressed.}}
\address{Department of Physics, Beijing Normal University, Beijing 100875, China}

\eads{\mailto{tuzc@bnu.edu.cn}}


\begin{abstract}
Typical heat engines exhibit a kind of homotypy: The heat exchanges between a cyclic heat engine and its two heat reservoirs abide by the same function type; the forward and backward flows of an autonomous heat engine also conform to the same function type. This homotypy mathematically reflects in the existence of hidden symmetries for heat engines. The heat exchanges between the cyclic heat engine and its two reservoirs are dual under the joint transformation of parity inversion and time-reversal operation. Similarly, the forward and backward flows in the autonomous heat engine are also dual under the parity inversion. With the consideration of these hidden symmetries, we derive a generic nonlinear constitutive relation up to the quadratic order for tight-coupling cyclic heat engines and that for tight-coupling autonomous heat engines, respectively.

\end{abstract}
\maketitle

\section{Introduction}

As an important source of power, heat engines are crucial to our human activities. It is necessary to investigate their energetics in our times of resource shortage. The classical equilibrium thermodynamics provides a powerful tool to investigate the ideal heat engines consisting of reversible processes. The concept of Cannot efficiency is one of the cornerstones of thermodynamics. It serves as the upper bound for efficiencies of heat engines working between two heat reservoirs.

Cannot efficiency is achieved only for heat engines operating in equilibrium or quasi-static states, which operate infinitely slow and yield vanishing power output. The practical heat engines usually complete thermodynamic cycles in finite periods or operating at finite net rates, which give rise to the development of finite-time thermodynamics. The most elegant result on this topic is the efficiency at maximum power for the Curzon-Alhborn endoreversible heat engine~\cite{Chambadal57,Novikov57,Curzon1975}, which reads $\eta_{CA}=1-\sqrt{T_{c}/T_{h}}$ with $T_{h}$ and $T_{c}$ being the temperatures of the hot reservoir and the cold one, respectively. This result has attracted much attention from physicists and engineers~\cite{VandenBroeck2013JCP,WangJH2012PRE031145,WangJH2012PRE051112,ChenJC2013PRE,WangJH2013PRE042119,QuanHT2014PRE,TuSheng14arxiv,Seifert2013PRL,Seifert2013NJP,Berry1977,devos85,Berry1985,Chen1989,ChenJC94,Bejan96,dcisbj2007,GaveauPRL10,Espositopre2010,Esposito2010,wangtu2012EPL,wangtu2012PRE,Schmiedl2008,Tu2008,Esposito2009EPL,vdbrk2005,Esposito2009PRL,ApertetPRE13,Izumida2009PRE,Izumida2014PRL,Izumida2012,TuShengJPA13,TuShengPRE14,Nakagawa06epl,VandenBroeck2012PRE}. The previous researches reveal that the Curzon-Alhborn efficiency ($\eta_{CA}$) is also recovered, or at least approximately recovered, in a lot of heat engines such as the stochastic heat engine \cite{Schmiedl2008}, the Feynman ratchet \cite{Tu2008}, the single-level quantum dot engine \cite{Esposito2009EPL}, and the symmetric low-dissipation heat engine \cite{Esposito2010}.

The connection between finite-time thermodynamics and linear irreversible thermodynamics was proposed by Van den Broeck~\cite{vdbrk2005}. The constitutive relation, which is defined as the relation between the generalized thermodynamic fluxes and forces, is one of the central formulas in irreversible thermodynamics. The constitutive relation is assumed to be linear and restricted by the Onsager reciprocal relation \cite{1Onsager1931} in linear irreversible thermodynamics. Using the linear constitutive relation, Van den Broeck found that the efficiency at maximum power for tight-coupling heat engines is half of the Carnot efficiency. However, we observed that many heat engines, such as the Curzon-Alhborn endoreversible heat engine~\cite{Curzon1975}, the Feynman ratchet as a heat engine~\cite{Tu2008} and the single-level quantum dot heat engine~\cite{Esposito2009EPL}, can not be precisely described by the constitutive relation for linear response. There exist higher order terms in the relations between generalized thermodynamic fluxes and forces for these heat engines. This conflict might lead to inappropriate results when we try to investigate the energetics of heat engines in a higher precision such as the universal efficiency at maximum power up to the quadratic order~\cite{Esposito2009PRL,TuShengPRE14,TuSheng14arxiv}. Thus, it is an urgent task for us to seek a generic expression of nonlinear constitutive relations for heat engines.

On the other hand, based on our early researches we noticed that typical heat engines exhibit a kind of homotypy, which had not been touched in previous literature. We observed that the heat exchanges between a cyclic heat engine and its two reservoirs abide by the same function type; and that the forward and backward flows for an autonomous heat engine also conform to the same function type. This homotypy mathematically reflects in the existence of hidden symmetries: The heat exchanges between the cyclic heat engine and its two reservoirs are dual under the joint transformation of parity inversion and time-reversal operation [see $\mathcal{PT}$-symmetry (\ref{Eq-homo-cyclicPT}) in Sec.~\ref{Sec-homocyclic-PT}]; the forward and backward flows in the autonomous heat engine are also dual under the parity inversion [see $\mathcal{P}$-symmetry (\ref{Eq-homo-autoP}) in Sec.~\ref{Sec-homoauto-P}]. It is still unclear what constraints will these hidden symmetries impose on the expressions of nonlinear constitutive relations for heat engines.

In this work, we first revisit the generic model for tight-coupling heat engines proposed in our previous work~\cite{TuShengPRE14}. Then we will investigate the hidden symmetries in heat engines, based on which we derive the nonlinear constitutive relations up to the quadratic order for cyclic heat engines and autonomous heat engines, respectively. These results are also confirmed by typical models of heat engines.

\section{Generic model\label{Sec-generic_model}}

In a conventional setup of heat engine, the engine absorbs heat $\dot{Q}_{h}$ from a hot reservoir at temperature $T_{h}$ and releases heat $\dot{Q}_{c}$ into a cold reservoir at temperature $T_{c}$ per unit time. Simultaneously, it outputs power $\dot{W}$ against an external load, which may be further expressed as
\begin{equation} \dot{W}=\dot{Q}_{h}-\dot{Q}_{c}, \label{Eq-model-conservation}\end{equation}
according to the energy conservation.

For the engine operating in a finite period or at finite rate rather than in a quasi-static state, the contribution of the interactions between the engine and the reservoirs should not be ignored \cite{TuShengPRE14}. We introduce two nonnegative weighted parameters $s_{h}$ and $s_{c}$ satisfying $s_{h}+s_{c}=1$ to represent the degree of asymmetry of the interactions between the engine and the reservoirs. With these parameters, the weighted thermal flux $J_{t}$ and the weighted reciprocal of temperature $\beta$ may be defined as
\begin{equation}J_{t}\equiv s_{h}\dot{Q}_{c}+s_{c}\dot{Q}_{h},\label{Eq-model-Jt}\end{equation}
and
\begin{equation}\beta\equiv s_{h}/T_{h}+s_{c}/T_{c},\label{Eq-model-beta}\end{equation}
respectively. The values of $s_{h}$ and $s_{c}$ depend on specific models. In particular, $s_{h}=s_{c}=1/2$ indicates a special situation in which the engine is symmetrically interacts with the two reservoirs.

\begin{figure}[htp!]
\centerline{\includegraphics[width=9cm]{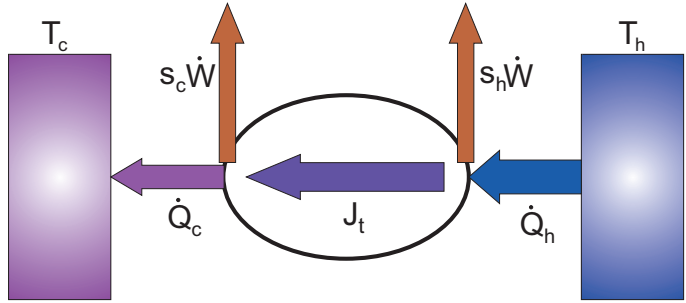}} \caption{\label{Fig1}Refined generic model of a tight-coupling heat engine (reproduced according to~\cite{TuShengPRE14}).}
\end{figure}

From (\ref{Eq-model-conservation}) and (\ref{Eq-model-Jt}), we obtain the heat fluxes
\begin{equation} \dot{Q}_{h}=J_{t}+s_{h}\dot{W},~~\mathrm{and}~~  \dot{Q}_{c}=J_{t}-s_{c}\dot{W}.\label{Eq-model-QcQh}\end{equation}
Based on these relations, we obtain a refined generic model as shown in Fig.~\ref{Fig1}. In this new physical picture, the engine absorbs heat $\dot{Q}_{h}$ from the hot reservoir per unit time while an amount of heat $s_{h}\dot{W}$ is converted into work due to the interaction between the engine and the hot reservoir. A thermal flux $J_{t}$ flows through the engine. Then the engine releases heat $\dot{Q}_{c}$ into the cold reservoir per unit time while an amount of heat $s_{c}\dot{W}$ is converted into work due to the interaction between the engine and the cold reservoir. The influence of the relative strength of the interactions between the engine and the reservoirs is explicitly included in this picture.

The generalized thermal force conjugated to $J_{t}$ may be defined as
\begin{equation} X_{t}\equiv 1/T_{c}-1/T_{h}.\label{Eq-model-Xt}\end{equation}
The definitions of generalized mechanical flux $J_{m}$ and mechanical force $X_{m}$ depend on the type of heat engines. For a cyclic heat engine, they may be defined as
\begin{equation} J_{m}\equiv 1/t_{0} \textmd{~~and~~} X_{m}\equiv -\beta W, \label{Eq-model-cyclicJmXm}\end{equation}
respectively, where $t_{0}$ is the period for completing a whole thermodynamic cycle. It should be noticed that the sign of $t_{0}$ is of physical meaning. $t_{0}$ takes a positive sign when the thermodynamic cycle operates as a heat engine, while it takes a negative sign when the thermodynamics cycle operates as an refrigerator. $W$ is the work output in each cycle. For an autonomous heat engine operating in a steady state, $J_{m}$ and $X_{m}$ may be defined as
\begin{equation} J_{m}\equiv r \textmd{~~and~~} X_{m}\equiv -\beta w, \label{Eq-model-autoJmXm}\end{equation}
respectively, where $r$ is the net flow and $w$ denotes the elementary work in each mechanical step.

With definitions (\ref{Eq-model-Jt})--(\ref{Eq-model-autoJmXm}), it is easy to verify that the entropy production rate of the whole system may be written in a canonical form
\begin{equation} \sigma=\dot{Q}_{c}/T_{c}-\dot{Q}_{h}/T_{h}=J_{m}X_{m}+J_{t}X_{t}.\label{Eq-model-sigma}\end{equation}
In this paper, we focus on a tight-coupling heat engine, in which the heat-leakage vanishes so that the thermal flux is proportional to the mechanical flux:
\begin{equation}J_{t}=\xi J_{m} \label{Eq-model-linearrelation2}\end{equation}
where $\xi$ is named coupling strength. Its physical meaning may be interpreted as the elementary thermal energy flowing through a cyclic engine in each period or that flowing through an autonomous heat engine in each spatial step. Define affinity
\begin{equation} A=X_{m}+\xi X_{t}, \label{Eq-model-A}\end{equation}
which represents the competition between the generalized mechanical force and the generalized thermal force. Then, the entropy production rate (\ref{Eq-model-sigma}) may be further expressed as
\begin{equation} \sigma =J_{m}A.\label{Eq-model-entropyfinal}\end{equation}

\section{Cyclic heat engines\label{Sec-homocyclic}}
In this section, we will investigate the hidden symmetry in cyclic heat engines and its influence on the constitutive relation for nonlinear response.

\subsection{$\mathcal{PT}$-symmetry for cyclic heat engines\label{Sec-homocyclic-PT}}
By analyzing typical models of cyclic heat engines in the literature such as the low-dissipation heat engine\cite{Esposito2010}, the Curzon-Alhborn heat engine~\cite{Curzon1975}, and the revised Curzon-Alhborn heat engine~\cite{Chen1989}, we observed that these models exhibit a kind of homotypy---the expression of the heat absorbed from the hot reservoir and that released into the cold reservoir abide by the same function type. For example, in low-dissipation heat engine, the heat absorbed from the hot reservoir and that released into the cold reservoir can be expressed as $Q_{h} =T_{h}( \Delta S-\Sigma_{h}/\tau_{h})$ and $-Q_{c} =T_{c}( -\Delta S-\Sigma_{c}/\tau_{c})$, respectively, which are exactly of the same function type. Here we have not explained the meanings of physical quantities in these two equations. The details can be found in Ref.\cite{Esposito2010}.
The other examples such as the Curzon-Alhborn heat engine and the revised one are fully discussed in Secs.~\ref{Sec-homocyclic-CA} and \ref{Sec-homocyclic-RECA}. We remind the reader to note equations (\ref{Eq-homocyclic-CAQh}), (\ref{Eq-homocyclic-CAQc}), and (\ref{Eq-homocyclic-RECAheat}).

This homotypy mathematically reflects in the existence of a hidden symmetry for cyclic heat engines. That is, under the parity-time ($\mathcal{PT}$) transformation, the heat exchanges between the engine and its two reservoirs are dual. Here the parity inversion indicates interchanging parameters related to the hot reservoir (the quantities with subscript $h$) and those related to the cold reservoir (the quantities with subscript $c$). The time-reversal operation changes the sign of time. This hidden symmetry can be mathematically expressed as
\begin{equation} \mathcal{PT} Q_{h}=Q_{c},~\mathrm{and}~\mathcal{PT}Q_{c}=Q_{h}.\label{Eq-homo-cyclicPT}\end{equation}
The above $\mathcal{PT}$-symmetry will be confirmed by the typical model of cyclic heat engines shown in the Sec.~\ref{Sec-homocyclic-CA} and Sec.~\ref{Sec-homocyclic-RECA}. It is not hard to check this point according to the following equations (\ref{Eq-homocyclic-CAQhQc}) and (\ref{Eq-homocyclic-RECAheatexchange}).

\subsection{Constitutive relation for nonlinear response\label{Sec-homocyclic-relation}}
We consider a cyclic heat engine undergoes a thermodynamic cycle consisting of two ``isothermal" and two adiabatic processes. The word ``isothermal" merely indicates that the heat engine is in contact with a heat reservoir at constant temperature.
In the process of ``isothermal" expansion during time interval $t_h$, the engine absorbs heat $Q_h$ from the hot reservoir at temperature $T_h$. The variation of entropy in this process is denoted as $\Delta S$. On the contrary, in the process of ``isothermal" compression during time interval $t_c$, the engine releases heat $Q_c$ into the cold reservoir at temperature $T_c$. There is no heat exchange and entropy production in two adiabatic processes. Assume that the time for completing the adiabatic processes is negligible relative to $t_c$ and $t_h$. So the period of the whole cycle is $t_0=t_c+t_h$. The heat exchanges $Q_h$ and $Q_c$ can be expressed as
\begin{equation}Q_{h} =T_{h}( \Delta S-S_h^{ir}),\textmd{~~and~~}-Q_{c} =T_{c}( -\Delta S-S_c^{ir}),\label{Eq-homocyclic-heat}\end{equation}
where $S_h^{ir}$ (or $S_c^{ir}$) represents the irreversible entropy production in the ``isothermal" expansion (or compression) process.
This model of cyclic heat engines is of broad generality if the entropy production in adiabatic processes of the thermodynamic cycle can be neglected. Equation (\ref{Eq-homocyclic-heat}) shows that $Q_{h}$ and $Q_{c}$ indeed abide by the same function type.

The ``isothermal'' expansion (or compression) process may be regarded as reversible in the long-time limit $t_{h}\rightarrow \infty$ (or $t_{c}\rightarrow \infty$). That means, $S_h^{ir}$ (or $S_c^{ir}$) should be vanishing when $t_{h}\rightarrow \infty$ (or $t_{c}\rightarrow \infty$). If the entropy production is an analytical function, we may write out
\begin{equation}\hspace{-10mm}S_h^{ir}=\frac{\Gamma_h}{t_h}+\Lambda_h \left(\frac{\Gamma_h}{t_h}\right)^2+O\left(\frac{\Gamma_h}{t_h}\right)^{3}, \textmd{~and~} S_c^{ir}=\frac{\Gamma_c}{t_c}+\Lambda_c \left(\frac{\Gamma_c}{t_c}\right)^2+O\left(\frac{\Gamma_c}{t_c}\right)^{3},\label{Eq-homocyclic-Sir}\end{equation}
with the time-independent parameters $\Gamma_h$, $\Lambda_h$, $\Gamma_c$, and $\Lambda_c$. The parameter $\Lambda_{h}$ either depends merely on the detailed ``isothermal" expansion process, or equals to some constant independent of specific processes. Similarly, $\Lambda_{c}$ either depends merely on the detailed ``isothermal" compression process, or equals to some constant independent of specific processes.
Up to the first order, equation (\ref{Eq-homocyclic-Sir}) degenerates into the so-called low-dissipation assumption proposed in \cite{Esposito2010}.

Substituting (\ref{Eq-homocyclic-Sir}) into (\ref{Eq-homocyclic-heat}), we obtain the expressions of the heat exchanges
\begin{equation}\begin{array}{l}Q_{h}=T_{h}\Delta S-T_{h}\bar\Gamma_{h}/t_{0}-T_{h}\bar\Gamma_{h}^2\Lambda_{h}/t_{0}^{2}+O(1/t_{0}^{3}),\\
                                Q_{c}=T_{c}\Delta S+T_{c}\bar\Gamma_{c}/t_{0}+T_{c}\bar\Gamma_c^2\Lambda_{c}/t_{0}^{2}+O(1/t_{0}^{3}),\end{array}\label{Eq-homocyclic-heatexchanges}\end{equation}
with parameters $\bar\Gamma_{h}\equiv \Gamma_{h}t_{0}/t_h$ and $\bar\Gamma_{c}\equiv \Gamma_{c}t_{0}/t_c$.

The $\mathcal{PT}$-symmetry (\ref{Eq-homo-cyclicPT}) between $Q_{h}$ and $Q_{c}$, will impose a constraint on the coefficients $\Lambda_{h}$ and $\Lambda_{c}$ in (\ref{Eq-homocyclic-heatexchanges}).
Substituting (\ref{Eq-homocyclic-heatexchanges}) into (\ref{Eq-homo-cyclicPT}), we find the only choice is
\begin{equation}\Lambda_{h}= -\Lambda_{c}=-\Lambda,\label{Eq-homocyclic-Lambda}\end{equation}
where $\Lambda$ is a process-independent parameter, and of course independent of the weighted parameters $s_{h}$ and $s_{c}$ which represent the degree of asymmetry of interactions between the engine and the reservoirs. With (\ref{Eq-homocyclic-Lambda}), the heat fluxes between the engine and two reservoirs may be expressed as
\begin{equation}\begin{array}{l}\dot{Q}_{h}\equiv Q_{h}/t_{0}=T_{h}\Delta S/t_{0}-T_{h}\bar\Gamma_{h}/t_{0}^{2}+O(1/t_{0}^{3}),\\
                                \dot{Q}_{c}\equiv Q_{c}/t_{0}=T_{c}\Delta S/t_{0}+T_{c}\bar\Gamma_{c}/t_{0}^{2}+O(1/t_{0}^{3}).\end{array}\label{Eq-homocyclic-flowsgeneric}\end{equation}

Now, we will construct the mapping from this cyclic heat engine into the generic model for tight-coupling heat engines mentioned in Sec.~\ref{Sec-generic_model}. Considering (\ref{Eq-homocyclic-flowsgeneric}), the weighted thermal flux (\ref{Eq-model-Jt}) may be further expressed as $J_{t}=(s_{c}T_{h}+s_{h}T_{c})\Delta S/t_{0}-(s_{c}T_{h}\bar\Gamma_{h}-s_{h}T_{c}\bar\Gamma_{c})/t_{0}^{2}+O(1/t_{0}^{3})$. According to the physical meaning of $J_{t}$, which has been fully discussed in \cite{TuShengPRE14}, the quadratic order term of $1/t_{0}$ should be vanishing. Thus, we require $s_{c}T_{h}\bar\Gamma_{h}-s_{h}T_{c}\bar\Gamma_{c}=0$. Combining with $s_{h}+s_{c}=1$, we obtain the weighted parameters
\begin{equation}s_{h}=\frac{T_{h}\bar\Gamma_{h}}{T_{h}\bar\Gamma_{h}+T_{c}\bar\Gamma_{c}},~~\mathrm{and}~~s_{c}=\frac{T_{c}\bar\Gamma_{c}}{T_{h}\bar\Gamma_{h}+T_{c}\bar\Gamma_{c}}.\label{Eq-homocyclic-shsc}\end{equation}
With these weighted parameters and definition (\ref{Eq-model-cyclicJmXm}) for cyclic heat engines, we further derive the weighted thermal flux (\ref{Eq-model-Jt}) and the weighted reciprocal of temperature (\ref{Eq-model-beta}) as
\begin{equation}J_{t}=\xi J_{m}+O(J_{m}^{3}),\label{Eq-homocyclic-Jt}\end{equation}
and
\begin{equation}\beta =\frac{\bar\Gamma_{h}+\bar\Gamma_{c}}{T_{h}\bar\Gamma_{h}+T_{c}\bar\Gamma_{c}},\label{Eq-homocyclic-beta}\end{equation}
with the coupling strength $\xi\equiv T_{h}T_{c}\beta\Delta S$. In this work, we focus on the constitutive relation accurate up to the quadratic order. Thus the third and higher order terms of $J_{m}$ could be neglected. Within this scope, equation (\ref{Eq-homocyclic-Jt}) implies that $J_{t}$ is still tightly coupled with $J_{m}$. From Eqs.~(\ref{Eq-homocyclic-heat}), (\ref{Eq-homocyclic-Sir}), (\ref{Eq-homocyclic-Lambda}), (\ref{Eq-homocyclic-beta}) and definition $J_{m}\equiv 1/t_{0}$ , we obtain the expression of the generalized mechanical force as
\begin{equation}\hspace{-10mm}X_{m}=-(T_{h}-T_{c})\beta\Delta S+(\bar\Gamma_{h}+\bar\Gamma_{c})J_{m}-(T_{h}\bar\Gamma_{h}^{2}-T_{c}\bar\Gamma_{c}^{2})\beta\Lambda J_{m}^{2}+O(J_{m}^{3}).\label{Eq-homocyclic-Xm}\end{equation}
From the above equation we finally solve the generic nonlinear constitutive relation for tight-coupling cyclic heat engines
\begin{equation}J_{m}=\frac{1}{\bar\Gamma_{h}+\bar\Gamma_{c}}A\left[1+\Lambda (s_{h}-s_{c}) A\right]+O(A^{3},X_{m}^{3}),\label{Eq-homocyclic-Jm}\end{equation}
with the consideration of (\ref{Eq-model-A}), (\ref{Eq-homocyclic-shsc}), (\ref{Eq-homocyclic-beta}) and $\xi=T_{h}T_{c}\beta \Delta S$.

For the low-dissipation heat engine~\cite{Esposito2010}, the entropy production (\ref{Eq-homocyclic-Sir}) in the ``isothermal'' expansion (or ``isothermal'' compression) process merely contains the first order term of $\Gamma_{h}/t_{h}$ (or $\Gamma_{c}/t_{c}$). Thus, the parameters $\Lambda_{h}$, $\Lambda_{c}$ as well as $\Lambda$ vanish in this model. Then, equation (\ref{Eq-homocyclic-Jm}) degenerates into a linear constitutive relation, $J_{m}=A/(\bar \Gamma_{h}+\bar \Gamma_{c})$, which is consistent with the result derived in~\cite{TuShengPRE14}.

\subsection{Curzon-Ahlborn endoreversible heat engine\label{Sec-homocyclic-CA}}

The Curzon-Ahlborn endoreversible heat engine \cite{Curzon1975} undergoes a thermodynamic cycle consisting of two isothermal processes and two adiabatic processes. In the isothermal expansion process, the working substance is in contact with a hot reservoir at temperature $T_{h}$. Its effective temperature is assumed to be $T_{he}$ ($T_{he}<T_{h}$). During time interval $t_{h}$, an amount of heat $Q_{h}$ is transferred from the hot reservoir to the working substance according to the heat transfer law
\begin{equation}\label{eq-transQh}
Q_{h}=\kappa_{h}(T_{h}-T_{he})t_{h},\label{Eq-homocyclic-CAQh}\end{equation}
where $\kappa_{h}$ is the thermal conductivity in this process. The variation of entropy in this process is denoted by $\Delta S$. In the isothermal compression process, the working substance is in contact with a cold reservoir at temperature $T_{c}$. Its effective temperature is $T_{ce}$ ($T_{ce}>T_{c}$). During time interval $t_{c}$, an amount of heat $Q_{c}$ is transmitted from the working substance into the cold reservoir according to the heat transfer law
\begin{equation}\label{eq-transQc}-Q_{c}=\kappa_{c}(T_{c}-T_{ce})t_{c},\label{Eq-homocyclic-CAQc}\end{equation}
where $\kappa_{c}$ denotes the thermal conductivity in this process. Expressions (\ref{Eq-homocyclic-CAQh}) and (\ref{Eq-homocyclic-CAQc}) show that the laws of heat absorbed and that released by the engine indeed abide by the same function type. The heat exchanges and the entropy productions are vanishing in the two adiabatic processes. Besides, the durations of these two adiabatic process are assumed to be negligible comparing to $t_{c}$ and $t_{h}$. Thus the period ($t_{0}$) for completing the whole cycle is $t_{c}+t_{h}$.

According to the discussion in \cite{wangtu2012EPL}, the endoreversible assumption proposed by Curzon and Ahlborn \cite{Curzon1975} leads to
\begin{equation} \Delta S=Q_{h}/T_{he}=Q_{c}/T_{ce}. \label{Eq-homocyclic-CAendoassu}\end{equation}
Thus, equations (\ref{Eq-homocyclic-CAQh}) and (\ref{Eq-homocyclic-CAQc}) may be transformed into
\begin{equation}{Q}_{h}=\frac{T_{h}\Delta S}{1+\Delta S/\gamma_{h}t_0}, \textmd{~~and~~} {Q}_{c}=\frac{T_{c}\Delta S}{1-\Delta S/\gamma_{c}t_0},\label{Eq-homocyclic-CAQhQc}\end{equation}
respectively, with two parameters $\gamma_{h}\equiv\kappa_{h}t_{h}/t_{0}$ and $\gamma_{c}\equiv\kappa_{c}t_{c}/t_{0}$. It is easy to verify that the heat exchanges [equation (\ref{Eq-homocyclic-CAQhQc})] indeed satisfy the $\mathcal{PT}$-symmetry (\ref{Eq-homo-cyclicPT}) for cyclic heat engines.

Now, we will map this Curzon-Ahlborn heat engine into the generic model detailed in Sec.~\ref{Sec-generic_model} and derive the constitutive relation for nonlinear response. According to (\ref{Eq-homocyclic-CAQhQc}), the heat fluxes may be expressed into
\begin{equation}\begin{array}{l}\dot Q_{h}\equiv Q_{h}/t_{0}=T_{h}\Delta S/t_{0}-T_{h}\Delta S^{2}/\gamma_{h}t_{0}^{2}+O(1/t_{0}^{3}),\\
                \dot Q_{c}\equiv Q_{c}/t_{0}=T_{c}\Delta S/t_{0}+T_{c}\Delta S^{2}/\gamma_{c}t_{0}^{2}+O(1/t_{0}^{3}).\end{array}\label{Eq-homocyclic-CAfluxes}\end{equation}
From (\ref{Eq-model-Jt}), (\ref{Eq-homocyclic-CAfluxes}) and the definition $J_{m}\equiv 1/t_{0}$ for cyclic heat engines, we can derive the weighted thermal flux $J_{t}=(s_{c}T_{h}+s_{h}T_{c})\Delta SJ_m-(s_{c}T_{h}/\gamma _{h}-s_{h}T_{c}/\gamma _{c})\Delta S^{2}J_m^{2}+O(J_m^{3})$. Considering the physical meaning of $J_{t}$ discussed in~\cite{TuShengPRE14}, we require $s_{c}T_{h}/\gamma _{h}-s_{h}T_{c}/\gamma _{c}=0$. Combining with $s_{h}+s_{c}=1$, we derive the weighted parameters
\begin{equation} s_{h}=\frac{T_{h}\gamma_{c}}{T_{h}\gamma_{c}+T_{c}\gamma_{h}}, \textmd{~~and~~} s_{c}=\frac{T_{c}\gamma_{h}}{T_{h}\gamma_{c}+T_{c}\gamma_{h}}.\label{Eq-homocyclic-CAShSc}\end{equation}
Then the weighted reciprocal of temperature (\ref{Eq-model-beta}) and weighted thermal flux (\ref{Eq-model-Jt}) may be expressed as
\begin{equation}\beta=\frac{\gamma_{h}+\gamma_{c}}{T_{h}\gamma_{c}+T_{c}\gamma_{h}},\label{Eq-homocyclic-CAbeta}\end{equation}
and
\begin{equation}J_{t}=\xi J_{m}+O(J_{m}^{3}),\label{Eq-homocyclic-CAJt}\end{equation}
with $\xi \equiv T_{h}T_{c}\beta \Delta S$. As mentioned above, we just focus on the constitutive relation accurate up to the quadratic order. In this sense, $J_{t}$ is still tightly coupled with $J_{m}$.

Substituting (\ref{Eq-homocyclic-CAQhQc}) and (\ref{Eq-homocyclic-CAbeta}) into (\ref{Eq-model-cyclicJmXm}), we obtain the generalized mechanical force
\begin{eqnarray}X_{m}=&-&(T_{h}-T_{c})\beta\Delta S+(T_{h}/\gamma_{h}+T_{c}/\gamma_{c})\beta\Delta S^{2}J_{m}\nonumber\\
&-&(T_{h}/\gamma_{h}^{2}-T_{c}/\gamma_{c}^{2})\beta\Delta S^{3}J_{m}^{2}+O(J_{m}^{3}),\label{Eq-homocyclic-CAXm}\end{eqnarray}
from which we finally solved the constitutive relation for the Curzon-Ahlborn heat engine accurate up to the quadratic order
\begin{equation}J_{m}=\frac{\gamma_{c}\gamma_{h}}{(\gamma_{c}+\gamma_{h})\Delta S^{2}}A\left(1+\frac{s_{h}-s_{c}}{\Delta S}A\right)+O(A^{3},X_{m}^{3}),\label{Eq-homocyclic-CAJm}\end{equation}
with the consideration of (\ref{Eq-model-A}), (\ref{Eq-homocyclic-CAShSc}), (\ref{Eq-homocyclic-CAbeta}) and $\xi =T_{h}T_{c}\beta \Delta S$. It is obvious that the constitutive relation (\ref{Eq-homocyclic-CAJm}) for the Curzon-Alhborn heat engine is a specific form of the generic nonlinear constitutive relation (\ref{Eq-homocyclic-Jm}) for tight-coupling cyclic heat engines with model-dependent parameters $\Lambda=1/\Delta S$, $\bar{\Gamma}_h=\Delta S^2/\gamma_h$, and $\bar{\Gamma}_c=\Delta S^2/\gamma_c$.

\subsection{Revised Curzon-Ahlborn heat engine\label{Sec-homocyclic-RECA}}

The thermodynamic processes and definitions of physical quantities in the revised Curzon-Ahlborn endoreversible heat engine~\cite{Chen1989} are exactly the same as those in the original Curzon-Ahlborn heat engine mentioned in Sec.~\ref{Sec-homocyclic-CA} except for the heat transfer law in two isothermal processes. Here the law of heat exchanges is revised to
\begin{equation}Q_{h}=\kappa_{h} (T_{he}^{-1}-T_{h}^{-1})t_{h},~~~\mathrm{and}~~-Q_{c}=\kappa_{c} (T_{ce}^{-1}-T_{c}^{-1})t_{c}.\label{Eq-homocyclic-RECAheat}\end{equation}
The above equation shows that the heat exchanges in two isothermal processes still abide by the same function type.

Considering the endoreversible assumption (\ref{Eq-homocyclic-CAendoassu}), we transform (\ref{Eq-homocyclic-RECAheat}) into
\begin{equation}\hspace{-10mm}Q_{h}=\frac{\gamma_{h}t_{0}}{2T_{h}}\left(\sqrt{1+\frac{4\Delta ST_{h}^{2}}{\gamma_{h}t_{0}}}-1\right), \textmd{~~and~~} Q_{c}=\frac{\gamma_{c}t_{0}}{2T_{c}}\left(1-\sqrt{1-\frac{4\Delta ST_{c}^{2}}{\gamma_{c}t_{0}}}\right),\label{Eq-homocyclic-RECAheatexchange}\end{equation}
respectively, with the parameters $\gamma_{h}\equiv\kappa_{h}t_{h}/t_{0}$ and $\gamma_{c}\equiv\kappa_{c}t_{c}/t_{0}$. It is easy to verify that the heat exchanges [equation (\ref{Eq-homocyclic-RECAheatexchange})] indeed satisfy the $\mathcal{PT}$-symmetry (\ref{Eq-homo-cyclicPT}) for cyclic heat engines.

This revised Curzon-Alhborn heat engine may be mapped into the generic model. The mapping procedure is similar to that in Sec.~\ref{Sec-homocyclic-CA}. Then we obtain
\begin{equation}s_{h}=\frac{T_{h}^{3}\gamma_{c}}{T_{h}^{3}\gamma_{c}+T_{c}^{3}\gamma_{h}},~~s_{c}=\frac{T_{c}^{3}\gamma_{h}}{T_{h}^{3}\gamma_{c}+T_{c}^{3}\gamma_{h}};\label{Eq-homocyclic-RECAShSc}\end{equation}
\begin{equation}\beta=\frac{T_{h}^{2}\gamma_{c}+T_{c}^{2}\gamma_{h}}{T_{h}^{3}\gamma_{c}+T_{c}^{3}\gamma_{h}},\label{Eq-homocyclic-RECAbeta}\end{equation}
\begin{equation}J_{t}=T_{h}T_{c}\beta \Delta S J_{m}+O(J_{m}^{3})=\xi J_{m}+O(J_{m}^{3}),\label{Eq-homocyclic-RECAJt}\end{equation}
and the nonlinear constitutive relation accurate up to the quadratic order
\begin{equation}J_{m}=\frac{\gamma_{h}\gamma_{c}}{(T_{h}^{2}\gamma_{c}+T_{c}^{2}\gamma_{h})\Delta S^{2}}A\left[1+\frac{2(s_{h}-s_{c})}{\Delta S}A\right]+O(A^{3},X_{m}^{3}).\label{Eq-homocyclic-RECAJm}\end{equation}
Obviously, this nonlinear constitutive relation for the revised Curzon-Ahlborn heat engine is a special form of the generic constitutive relation (\ref{Eq-homocyclic-Jm}) for tight-coupling cyclic heat engines with model-dependent parameters $\Lambda =2/\Delta S$, $\bar{\Gamma}_h=T_h^2\Delta S^2/\gamma_h$, and $\bar{\Gamma}_c=T_c^2\Delta S^2/\gamma_c$.

\section{Autonomous heat engines\label{Sec-homoauto}}
In this section, we will investigate the hidden symmetry in autonomous heat engines and its influence on the constitutive relation for nonlinear response.

\subsection{$\mathcal{P}$-symmetry for autonomous heat engines\label{Sec-homoauto-P}}

By analyzing typical models of autonomous heat engines in the literature such as the Feynman ratchet as a heat engine \cite{Tu2008} and the single-level quantum dot heat engine \cite{Esposito2009EPL}, we found that the expressions of the forward and backward flows in an autonomous heat engine conform to the same function type. For example, the forward and backward flows in the Feynman ratchet as a heat engine~\cite{Tu2008} may be expressed as $R_{F} =r_{0}\mathrm{e}^{-(\epsilon +z\theta _{h})/T_h}$ and $R_{B}=r_{0}\mathrm{e}^{-(\epsilon -z\theta _{c})/T_c}$~\cite{TuShengPRE14}, respectively, where the detailed meanings of physical quantities will be fully explained in Sec.~\ref{Sec-homo-Feynman}. Obviously, the forward and backward flows conform to the same function type. This homotypy reflects in the existence of a hidden symmetry under the parity ($\mathcal{P}$) inversion discussed as below. The parity inversion indicates interchanging the quantities related to the hot reservoir (the quantities with subscript $h$) and the the quantities related to the cold reservoir (the quantities with subscript $c$), and simultaneously reversing the sign of the external load. Under the parity inversion, the expression of forward flow would turn into the expression of backward flow, and vice versa. This $\mathcal{P}$-symmetry may be mathematically expressed as
\begin{equation} \mathcal{P}R_{F}=R_{B},~~\mathrm{and}~~\mathcal{P}R_{B}=R_{F}.\label{Eq-homo-autoP}\end{equation}

On the other hand, the forward flow in an autonomous heat engine should merely rely on the intrinsic quantities related to the hot reservoir, such as $T_{h}$, the temperature of the hot reservoir, and $s_{h}$, the relative strength of interaction between the engine and the hot reservoir. The backward flow should merely rely on the intrinsic quantities related to the cold reservoir, such as $T_{c}$, the temperature of the cold reservoir, and $s_{c}$, the relative strength of interaction between the engine and the cold reservoir. Thus, combining with the $\mathcal{P}$-symmetry (\ref{Eq-homo-autoP}), the forward flow $R_{F}$ and backward flow $R_{B}$ may be formally expressed as
\begin{equation} R_{F}=\Phi (s_{h}X_{m},~1/T_{h}),~~\mathrm{and}~~R_{B}=\Phi (-s_{c}X_{m},~1/T_{c}),\label{Eq-homo-autoRfRb}\end{equation}
respectively, where $\Phi (x,~y)$ represents a function type with independent variables $x$ and $y$. The opposite signs before the terms related to $X_{m}$ in $R_{F}$ and $R_{B}$ originate from the fact that the external load changes its sign under the parity inversion. The $\mathcal{P}$-symmetry (\ref{Eq-homo-autoP}) as well as the formal expression (\ref{Eq-homo-autoRfRb}) of the flows will be confirmed by the examples of autonomous heat engines shown in Sec.~\ref{Sec-homo-Feynman} and Sec.~\ref{Sec-homoauto-quantum}.

\subsection{Constitutive relation for nonlinear response\label{Sec-homoauto-relation}}

\begin{figure}[htp!]
\centerline{\includegraphics[width=15cm]{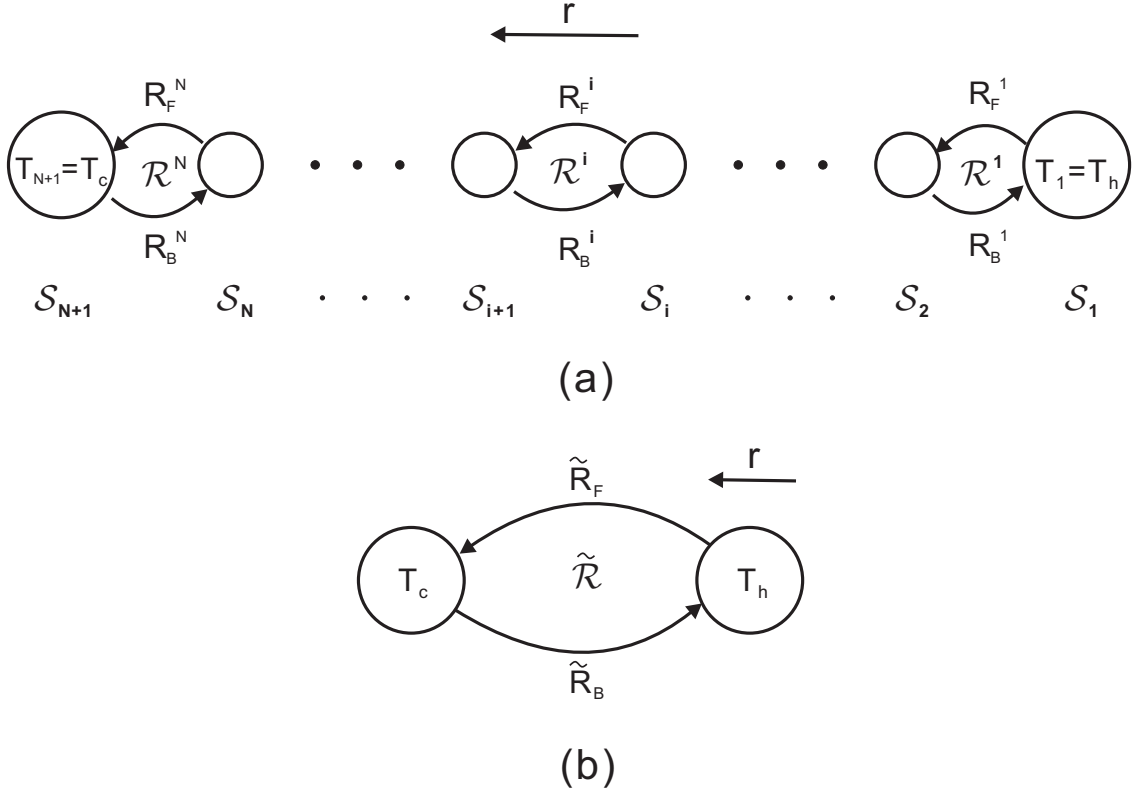}} \caption{\label{Fig2} (a) Schematic of a generic autonomous heat engine; (b) Schematic of reduced autonomous heat engine.}
\end{figure}

The schematic of a generic autonomous heat engine is depicted in Fig.~\ref{Fig2}(a). In this setup, the heat engine consist of $N+1$ separate states $\mathcal{S}_{n}$ $(n=1,2,\cdots,N+1)$, in which $\mathcal{S}_{1}$ denotes the hot reservoir at temperature $T_1=T_h$ and $\mathcal{S}_{N+1}$ denotes the cold reservoir at temperature $T_{N+1}=T_{c}$. The particles are transmitted sequentially from the hot reservoir $\mathcal{S}_{1}$ into the cold reservoir $\mathcal{S}_{N+1}$ through the $N-1$ intermediate states. Thus the whole transmission process $\mathcal{R}$ contains $N$ elementary transmission processes $\mathcal{R}^{i}$ $(i=1,2,\cdots,N)$ between the adjacent states as shown in Fig.~\ref{Fig2}(a). Each elementary process $\mathcal{R}^{i}$ is composed by a forward process with flow $R_{F}^{i}$ and a backward process with flow $R_{B}^{i}$. It should be noted that $R_{F}^{i}$ (or $R_{B}^{i}$) is the product of the pure forward (or backward) jumping rate in elementary process $\mathcal{R}^{i}$ and the occupation probability of state $\mathcal{S}_{i}$ (or $\mathcal{S}_{i+1}$). When the engine operates in the steady state, the relation
\begin{equation} R_{F}^{i}-R_{B}^{i}=r,\label{Eq-homoauto-r}\end{equation}
is satisfied in each elementary process $\mathcal{R}^{i}$ $(i=1,2,\cdots,N)$, where $r$ is the net flow of the whole transmission process $\mathcal{R}$. In this engine, along the particle flow $r$, the heat absorbed from the hot reservoir $\mathcal{S}_{1}$ is partly released into the cold reservoir $\mathcal{S}_{N+1}$ and partly outputted in the form of work against a globally generalized external force $X_{m}$.

It has been proved that the entropy production rate may be expressed as $\sigma =r  {\sum_{i=1}^{N}}\ln({R_{F}^{i}}/{R_{B}^{i}})$ for autonomous heat engines operating in a steady state~\cite{Lebowitz99,Seifert12rev}. Comparing it with (\ref{Eq-model-entropyfinal}), we obtain the expression of affinity
\begin{equation} A= \displaystyle{\sum_{i=1}^{N}}\ln\frac{R_{F}^{i}}{R_{B}^{i}},\label{Eq-homoauto-A}\end{equation}
by considering definition $J_{m}\equiv r$ for autonomous heat engines. Substituting (\ref{Eq-homoauto-r}) into (\ref{Eq-homoauto-A}), the affinity may be further expressed in terms of $R_{F}^{i}$ or $R_{B}^{i}$ as
\begin{equation}\begin{array}{l} A= \sum\limits_{i=1}^{N}\ln\frac{R_{F}^{i}}{R_{F}^{i}-r}=\sum\limits_{i=1}^{N}\frac{1}{R_{F}^{i}}r+\frac{1}{2}\sum\limits_{i=1}^{N}\frac{1}{{R_{F}^{i}}^{2}}r^{2}+O(r^{3}),\\
A= \sum\limits_{i=1}^{N}\ln\frac{R_{B}^{i}+r}{R_{B}^{i}}=\sum\limits_{i=1}^{N}\frac{1}{R_{B}^{i}}r-\frac{1}{2}\sum\limits_{i=1}^{N}\frac{1}{{R_{B}^{i}}^{2}}r^{2}+O(r^{3}),\end{array}\label{Eq-homoauto-ARfRb}\end{equation}
respectively, where $O(r^{3})$ denotes the third and higher order terms of $r$.

Assume that the forward and backward flows in all elementary transmission processes $\mathcal{R}^{i}$ $(i=1,2,\cdots,N)$ conform to the same function type. Focus on one elementary transmission process $\mathcal{R}^{i}$ between state $\mathcal{S}_i$ at effective temperature $T_i$ and state $\mathcal{S}_{i+1}$ at effective temperature $T_{i+1}$, which could be regarded as a minimal setup of autonomous engine operating between ``hot" reservoir $\mathcal{S}_i$ and ``cold" reservoir $\mathcal{S}_{i+1}$. According to (\ref{Eq-homo-autoRfRb}), $1/R_{F}^{i}$ and $1/R_{B}^{i}$ can be formally expressed as
\begin{equation}\begin{array}{l}1/R_{F}^{i}=1/\Phi (s_{h}^{i}X_{m}^{i},1/T_{i})=\phi_{i}+\phi_{i}^{\prime}s_{h}^{i}X_{m}^{i}+O({X_{m}^{i}}^{2}),\\
1/R_{B}^{i}=1/\Phi (-s_{c}^{i}X_{m}^{i},1/T_{i+1})=\phi_{i+1}-\phi_{i+1}^{\prime}s_{c}^{i}X_{m}^{i}+O({X_{m}^{i}}^{2}),\end{array}\label{Eq-homoauto-RfRbseries}\end{equation}
respectively,
where $s_{h}^{i}$ and $s_{c}^{i}$ (satisfying $s_{h}^{i}+s_{c}^{i}=1$) are weighted parameters in this minimal setup of heat engine, which denote the asymmetry degree of interaction in elementary transmission process $\mathcal{R}^{i}$. $X_{m}^{i}$ satisfying $\sum_{i=1}^{N}X_{m}^{i}=X_{m}$, denotes the locally generalized external force in process $\mathcal{R}^{i}$. Then we can define the weighted parameters for the whole autonomous heat engine as
\begin{equation}s_{h}=\frac{1}{X_{m}}\sum\limits_{i=1}^{N}s_{h}^{i}X_{m}^{i},~~\mathrm{and}~~s_{c}=\frac{1}{X_{m}}\sum\limits_{i=1}^{N}s_{c}^{i}X_{m}^{i}.\label{Eq-homoauto-ShSc}\end{equation}
In (\ref{Eq-homoauto-RfRbseries}) we have defined the parameters $\phi_{i}$ and $\phi_{i}^{\prime}$ as
\begin{equation}\phi_{i}\equiv \left.\frac{1}{\Phi}\right|_{(x=0,y=1/T_{i})},~~\mathrm{and}~~\phi_{i}^{\prime}\equiv \left.\frac{\partial(1/\Phi)}{\partial x}\right|_{(x=0,y=1/T_{i})},\label{Eq-homoauto-Phi0Phi1}\end{equation}
If we further introduce an average quantity of $\phi _{i}^{\prime}~(i=1,2,\cdots,N+1)$ as
\begin{equation}\overline{\phi^{\prime}}=\frac{1}{N+1}\sum_{i=1}^{N+1}\phi_{i}^{\prime},\label{eq-phiprime}\end{equation}
it is easy to verify
\begin{equation} \phi_{i}^{\prime}-\overline{\phi^{\prime}}=O(X_{t})~~(1\leq i\leq N+1),\label{Eq-homoauto-deltaPhi1}\end{equation}
and
\begin{equation} \phi_{i}-\phi_{j}=O(X_{t})~~(1\leq i,j\leq N+1),\label{Eq-homoauto-deltaPhi0}\end{equation}
from definitions (\ref{Eq-model-Xt}), (\ref{Eq-homoauto-Phi0Phi1}) and $1/T_{h}\leq 1/T_{i}\leq 1/T_{c}$.

Substituting (\ref{Eq-homoauto-RfRbseries})--(\ref{Eq-homoauto-deltaPhi0}) into (\ref{Eq-homoauto-ARfRb}), we obtain the expression of $A$ as
\begin{eqnarray} A&=&\frac{1}{2}\left(\sum\limits_{i=1}^{N}\ln \frac{R_{F}^{i}}{R_{F}^{i}-r}+\sum\limits_{i=1}^{N}\ln \frac{R_{B}^{i}+r}{R_{B}^{i}}\right)\nonumber\\
&=&\left(\sum_{i=1}^{N}\frac{\phi_{i}+\phi_{i+1}}{2}\right)r+\frac{\overline{\phi^{\prime}}}{2}(s_{h}-s_{c})X_{m}r+O(X_{m}^{3},r^{3}).\label{Eq-homoauto-Afinal}\end{eqnarray}
From (\ref{Eq-homoauto-Afinal}) and definition $J_{m}\equiv r$, we can solve the constitutive relation for nonlinear response
\begin{equation}J_{m}=LA\left[1-\frac{L\overline{\phi^{\prime}}}{2}(s_{h}-s_{c})X_{m}\right]+O(A^{3},X_{m}^{3}),\label{Eq-homoauto-Jm}\end{equation}
with the parameter $L=[\sum_{i=1}^{N}(\phi_{i}+\phi_{i+1})/2]^{-1}$.

\subsection{Feynman ratchet\label{Sec-homo-Feynman}}

\begin{figure}[htp!]
\centerline{\includegraphics[width=10cm]{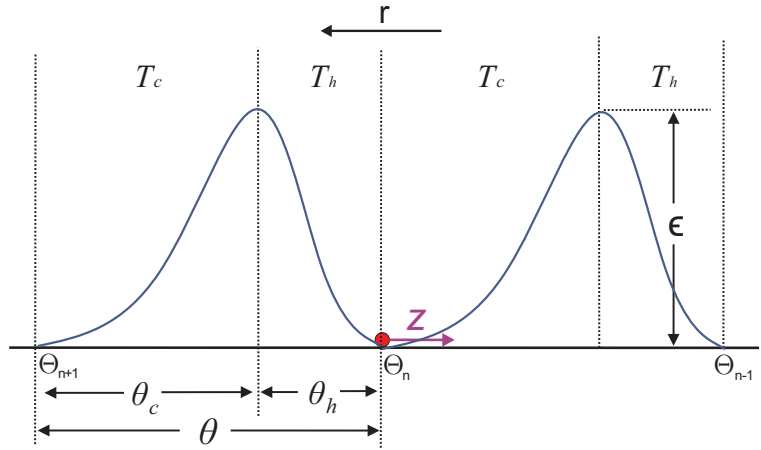}} \caption{\label{Fig3} Schematic digram of Feynman ratchet as a heat engine.}
\end{figure}

The Feynman ratchet \cite{1Feynmanbook,1Buttiker,1Landauer,1Seifert09PRE,1Apertet2014PRE} may be regarded as a Brownian particle walking in a periodic potential with a fixed step size $\theta$.
As depicted in Fig.~\ref{Fig3}, the Brownian particle is in contact with a hot reservoir at temperature $T_h$ in the right side of each energy barrier while it is in contact with a cold reservoir at temperature $T_c$ in the left side of each barrier. The particle moves across each barrier from right to left and outputs work against an external load $z$. The height of energy barrier is $\epsilon$. The width of potential in the left or right side of the barrier is denoted by $\theta_{c}$ or $\theta_{h}=\theta-\theta_{c}$, respectively. In the steady state and under the overdamping condition, according to the Arrhenius law, the forward and backward flows can be respectively expressed as
\begin{equation}\label{Eq-homoauto-FeynmanRfRb1}
R_{F} =r_{0}\mathrm{e}^{-(\epsilon +z\theta _{h})/T_h}, \mathrm{~~and~~} R_{B}=r_{0}\mathrm{e}^{-(\epsilon -z\theta _{c})/T_c},\end{equation} where $r_{0}$ represents the bare rate constant with dimension of time$^{-1}$. According to (\ref{Eq-homoauto-FeynmanRfRb1}), the forward and backward flows in the Feynman ratchet indeed conform to the same function type. The net flow in the Feynman ratchet may be defined as $r\equiv R_{F}-R_{B}$.

In each forward step, the particle absorbs heat $\epsilon +z\theta_{h}$ from the hot reservoir and releases heat $\epsilon -z\theta_{c}$ into the cold reservoir, while outputs work $w=z\theta$ against the external load. The energy conversion in each backward step is exactly opposite to that in the forward step. Thus, the heat absorbed from the hot reservoir and that released into the cold reservoir per unit time, as well as the power output may be further expressed as
\begin{equation}\begin{array}{l}\dot{Q}_{h}=(\epsilon+z\theta_{h})r=\epsilon r+(\theta_{h}/\theta )\dot{W},\\
                             \dot{Q}_{c}=(\epsilon-z\theta_{c})r=\epsilon r-(\theta_{c}/\theta )\dot{W},\end{array}\label{Eq-homoauto-FeynmanQhQc}\end{equation}
and
\begin{equation} \dot{W}=z\theta r,\label{Eq-homoauto-Feynmanpower}\end{equation}
respectively. Comparing (\ref{Eq-homoauto-FeynmanQhQc}) with (\ref{Eq-model-QcQh}), we can straightforwardly write out the weighted parameters
\begin{equation} s_{h}=\theta_{h}/\theta ,~~\mathrm{and}~~s_{c}=\theta_{c}/\theta .\label{Eq-homoauto-FeynmanShSc}\end{equation}
Substituting (\ref{Eq-homoauto-FeynmanShSc}) into (\ref{Eq-model-Jt}) and (\ref{Eq-model-beta}), we derive
\begin{equation}J_{t}=\epsilon r,~~\mathrm{and}~~\beta=(\theta_{h}/T_{h}+\theta_{c}/T_{c})/\theta ,\label{Eq-homoauto-FeynmanJtbeta}\end{equation}
from which we derive $\xi=\epsilon$ by using (\ref{Eq-model-linearrelation2}) and definition $J_{m}=r$ for autonomous heat engines.
From (\ref{Eq-model-autoJmXm}) and (\ref{Eq-homoauto-Feynmanpower}), we obtain the expression of generalized mechanical force
\begin{equation}X_{m}=-\beta z\theta.\label{Eq-homoauto-FeynmanXm}\end{equation}
Considering (\ref{Eq-homoauto-FeynmanShSc})--(\ref{Eq-homoauto-FeynmanXm}), we transform (\ref{Eq-homoauto-FeynmanRfRb1}) into
\begin{equation}R_{F}=r_{0}e^{-(\epsilon-s_{h}X_{m}/\beta)/T_{h}},~~\mathrm{and}~~R_{B}=r_{0}e^{-(\epsilon+s_{c}X_{m}/\beta)/T_{c}}.\label{Eq-homoauto-FeynmanRfRb2}\end{equation}
It is easy to verify that the forward and backward flows in (\ref{Eq-homoauto-FeynmanRfRb2}) indeed satisfy the $\mathcal{P}$--symmetry (\ref{Eq-homo-autoP}) and conform to the formal expression (\ref{Eq-homo-autoRfRb}).

Substituting (\ref{Eq-model-Xt}), (\ref{Eq-model-A}) , (\ref{Eq-homoauto-FeynmanXm}) and (\ref{Eq-homoauto-FeynmanRfRb2}) into $J_m \equiv  r= R_{F}-R_{B}$, we obtain the nonlinear constitutive relation up to the quadratic order as
\begin{equation} J_{m}=r_{0}\textrm{e}^{-\bar{\beta}\epsilon}A\left[1+\frac{1}{2}(s_{h}-s_{c})X_{m}\right] +O(A^{3}, X_{m}^{3}),\label{Eq-homoauto-FeynmanJm}\end{equation}
where $\bar{\beta}=(1/T_{h}+1/T_{c})/2$. Obviously, equation (\ref{Eq-homoauto-FeynmanJm}) is a specific form of the generic nonlinear constitutive relation (\ref{Eq-homoauto-Jm}) for autonomous heat engines.

From another point of view, we can also derive the constitutive relation (\ref{Eq-homoauto-FeynmanJm}) from (\ref{Eq-homoauto-Jm}). Comparing the forward and backward flows of the Feynman ratchet [in equation (\ref{Eq-homoauto-FeynmanRfRb2})] with the formal expression (\ref{Eq-homo-autoRfRb}) of forward and backward flows for autonomous heat engines, we could derive the specific form of function $\Phi (x,y)$ for the Feynman ratchet as
\begin{equation} \Phi (x,y)=r_{0}e^{xy/\beta -\epsilon y}.\label{Eq-homoauto-FeynmanPhi}\end{equation}
Because the Feynman ratchet may be regarded as a setup consisting of only one elementary transmission process between two states (i.e., two reservoirs), we use subscript $h$ to denote the quantities related to the hot reservoir and subscript $c$ to denote the quantities related to the cold reservoir. Then, substituting (\ref{Eq-homoauto-FeynmanPhi}) into (\ref{Eq-homoauto-Phi0Phi1}) and (\ref{eq-phiprime}), we obtain
\begin{equation}\phi_{h}=\frac{1}{r_{0}}e^{\epsilon/T_{h}},~~\phi_{c}=\frac{1}{r_{0}}e^{\epsilon/T_{c}},\label{Eq-homoauto-Feynmanphi0}\end{equation}
and
\begin{equation}\overline{\phi^{\prime}}=\frac{1}{2}({\phi}_{h}^{\prime}+{\phi}_{c}^{\prime})=-\frac{1}{2r_{0}\beta}\left(\frac{1}{T_{h}}e^{\epsilon/T_{h}}+\frac{1}{T_{c}}e^{\epsilon/T_{c}}\right).\label{Eq-homoauto-Feynmanphi1}\end{equation}
Finally, substituting (\ref{Eq-homoauto-Feynmanphi0}) and (\ref{Eq-homoauto-Feynmanphi1}) into the generic nonlinear constitutive relation (\ref{Eq-homoauto-Jm}) for autonomous heat engines, we can achieve the nonlinear constitutive relation (\ref{Eq-homoauto-FeynmanJm}) for the Feynman ratchet again.

\subsection{Single-level quantum dot heat engine\label{Sec-homoauto-quantum}}

A single-level quantum dot heat engine~\cite{Esposito2009EPL} consists of three parts: a hot lead at temperature $T_{h}$ and chemical potential $\mu_{h}$; a cold lead at temperature $T_{c}$ ($T_{c}<T_{h}$) and chemical potential $\mu_{c}$ ($\mu_{c}>\mu_{h}$); a single-level quantum dot with energy level $\varepsilon$ ($\varepsilon>\mu_{c}$), which located between the two leads. In the forward process, an electron jumps from the hot lead to the cold one via the quantum dot. The electron absorbs heat $q_{h}\equiv \varepsilon-\mu_{h}$ from the hot lead and releases heat $q_{c}\equiv\varepsilon-\mu_{c}$ into the cold one, and simultaneously outputs work $w\equiv\mu_{c}-\mu_{h}$. In the backward process, the energy conversion exactly opposites to that in the forward process. When the engine operates in a steady state, the overall forward and backward electronic flows may be expressed as~\cite{Esposito2009EPL}:
\begin{equation}r_{F}=\frac{ \alpha}{e^{(\varepsilon-\mu_{h})/T_{h}}+1},~~\mathrm{and}~~r_{B}=\frac{\alpha }{e^{(\varepsilon-\mu_{c})/T_{c}}+1},\label{Eq-homoauto-quantumRfRb}\end{equation}
respectively, with the coefficient $\alpha$. The above equation indicates the overall forward and backward flows of this single-level quantum dot engine indeed conform to the same function type. Then the net flow from the hot reservoir into the cold one can be written as
\begin{equation}J_m\equiv r_{F}-r_{B}=\alpha\left[\frac{1}{e^{(\varepsilon-\mu_{h})/T_{h}}+1}-\frac{1}{e^{(\varepsilon-\mu_{c})/T_{c}}+1}\right].\label{Eq-homoauto-quantumJm1}\end{equation}
The heat absorbed from the hot reservoir and that released into the cold reservoir per unit time, as well as the power output could be further expressed as
\begin{equation}\dot{Q}_{h}=(\varepsilon-\mu_{h})J_m,~\dot{Q}_{c}=(\varepsilon-\mu_{c})J_m,\label{Eq-homoauto-quantumfluxes}\end{equation}
and
\begin{equation}\dot{W}=(\mu_{c}-\mu_{h})J_m,\label{Eq-homoauto-quantumP}\end{equation}
respectively. As the engine operates in the steady state, the quantum dot is assumed to be locally in equilibrium. Thus we may introduce $\mu$ ($\mu_{h}\leq\mu\leq\mu_{c}$) as the effective chemical potential of the quantum dot. Then we can transform (\ref{Eq-homoauto-quantumfluxes}) into
\begin{equation}\dot{Q}_{h}=(\varepsilon-\mu)J_m+\left(\frac{\mu-\mu_{h}}{\mu_{c}-\mu_{h}}\right)\dot{W},~~\dot{Q}_{c}=(\varepsilon-\mu)J_m-\left(\frac{\mu_{c}-\mu}{\mu_{c}-\mu_{h}}\right)\dot{W}.\label{Eq-homoauto-quantumfluxes2}\end{equation}
Comparing (\ref{Eq-homoauto-quantumfluxes2}) with (\ref{Eq-model-QcQh}), we have
\begin{equation}s_{h}=\frac{\mu-\mu_{h}}{\mu_{c}-\mu_{h}},~s_{c}=\frac{\mu_{c}-\mu}{\mu_{c}-\mu_{h}},\label{Eq-homoauto-quantumShSc}\end{equation}
\begin{equation}\mu=s_{c}\mu_{h}+s_{h}\mu_{c},\label{Eq-homoauto-quantummu}\end{equation}
and
\begin{equation}J_{t}=\xi J_{m},\label{Eq-homoauto-quantumJt}\end{equation}
with the coupling strength $\xi =\varepsilon-\mu$. Substituting (\ref{Eq-homoauto-quantumShSc}) into (\ref{Eq-model-beta}) and (\ref{Eq-model-autoJmXm}), we derive the expressions of weighted reciprocal of temperature and generalized mechanical force as
\begin{equation}\beta=s_{h}/T_{h}+s_{c}/T_{c}=\frac{1}{\mu_{c}-\mu_{h}}\left(\frac{\mu-\mu_{h}}{T_{h}}+\frac{\mu_{c}-\mu}{T_{c}}\right),\label{Eq-homoauto-quantumbeta}\end{equation}
and
\begin{equation} X_{m}=-\beta w=(\mu_{h}-\mu)/T_{h}-(\mu_{c}-\mu)/T_{c}.\label{Eq-homoauto-quantumXm}\end{equation}
With definitions (\ref{Eq-homoauto-quantumShSc}) and (\ref{Eq-homoauto-quantumXm}), the overall forward and backward flows in (\ref{Eq-homoauto-quantumRfRb}) could be transformed into
\begin{equation}r_{F}=\frac{\alpha}{e^{\left[(\varepsilon-\mu)-s_{h}X_{m}/\beta \right]/T_{h}}+1},~~\mathrm{and}~~r_{B}=\frac{\alpha}{e^{\left[(\varepsilon-\mu)+s_{c}X_{m}/\beta \right]/T_{c}}+1}.\label{Eq-homoauto-quantumRfRb2}\end{equation}
Obviously the above equation satisfies $\mathcal{P}$-symmetry (\ref{Eq-homo-autoP}) and abides by the formal expression (\ref{Eq-homo-autoRfRb}) of the forward and backward flows. Besides, the specific form of function $\Phi (x,y)$ for single-level quantum dot engine can be written as $\Phi (x,y)=\alpha /\left[e^{(\varepsilon-\mu)y-xy/\beta}+1\right]$.

Substituting (\ref{Eq-homoauto-quantumShSc})--(\ref{Eq-homoauto-quantumXm}) and (\ref{Eq-model-A}) into (\ref{Eq-homoauto-quantumJm1}), we finally obtain the constitutive relation for nonlinear response as
\begin{equation} J_{m}=\frac{\alpha}{4\cosh^{2}(\bar{\beta}\xi/2)}A\left[1+\frac{\tanh(\bar{\beta}\xi/2)}{2}(s_{h}-s_{c})X_{m}\right]+O(A^{3}, X_{m}^{3}),\label{Eq-homoauto-quantumJm2}\end{equation}
with $\bar{\beta}=(T_{c}^{-1}+T_{h}^{-1})/2$. Obviously, the above equation is a specific form of the nonlinear constitutive relation (\ref{Eq-homoauto-Jm}) for tight-coupling autonomous heat engines.

\subsection{Reduction of autonomous heat engines\label{Sec-homoauto-simply}}

In the steady state, the entropy production rate of the multi-step autonomous heat engine shown in Fig.~\ref{Fig2}(a) may be expressed as $\sigma =r \sum_{i=1}^{N}\ln(R_{F}^{i}/R_{B}^{i})$, where $r\equiv R_{F}^{i}-R_{B}^{i}$ is satisfied in each elementary transmission process $\mathcal{R}^{i}$ ($i=1,2,\cdots,N$). From the perspective of entropy production, if we define an effective transmission process $\widetilde{\mathcal{R}}$ with forward rate $\widetilde{R}_{F}$ and backward rate $\widetilde{R}_{B}$, which satisfy
\begin{equation}\widetilde{R}_{F}-\widetilde{R}_{B}=r,~~\mathrm{and}~~\ln \frac{\widetilde{R}_{F}}{\widetilde{R}_{B}}=\sum\limits_{i=1}^{N}\ln \frac{R_{F}^{i}}{R_{B}^{i}}=A,\label{Eq-homoauto-simiplydefine}\end{equation}
the multi-step model of autonomous heat engines shown in Fig.~\ref{Fig2}(a) may be reduced to an effective model shown in Fig.~\ref{Fig2}(b) with effective forward flow $\widetilde{R}_{F}$ and backward flow $\widetilde{R}_{B}$.
Then, from definition (\ref{Eq-homoauto-simiplydefine}), we can derive the expressions of $1/\widetilde{R}_{F}$ and $1/\widetilde{R}_{B}$ as
\begin{equation}\begin{array}{l}\frac{1}{\widetilde{R}_{F}}=\sum\limits_{i=1}^{N}\frac{1}{R_{F}^{i}} -\frac{1}{2}\sum\limits_{i\neq j}^{N}\frac{1}{R_{F}^{i}R_{F}^{j}}r+\frac{1}{6}\sum\limits_{i\neq j\neq k}^{N}\frac{1}{R_{F}^{i}R_{F}^{j}R_{F}^{k}}r^{2}+O(r^{3}),\\
\frac{1}{\widetilde{R}_{B}}=\sum\limits_{i=1}^{N}\frac{1}{R_{B}^{i}} +\frac{1}{2}\sum\limits_{i\neq j}^{N}\frac{1}{R_{B}^{i}R_{B}^{j}}r+\frac{1}{6}\sum\limits_{i\neq j\neq k}^{N}\frac{1}{R_{B}^{i}R_{B}^{j}R_{B}^{k}}r^{2}+O(r^{3}).\end{array}\label{Eq-homoauto-simplyRfRb}\end{equation}
From (\ref{Eq-homoauto-A}) and (\ref{Eq-homoauto-simiplydefine}), the affinity may be expressed as $A=\ln [ {\widetilde{R}_{F}}/({\widetilde{R}_{F}-r})]=r/{\widetilde{R}_{F}}+r^2/{2\widetilde{R}_{F}^{2}}+O(r^{3})$ or $A=\ln [(\widetilde{R}_{B}+r)/\widetilde{R}_{B}]=r/{\widetilde{R}_{B}}-r^2/{2\widetilde{R}_{B}^{2}}+O(r^{3})$. From these expressions, we may derive the affinity and entropy production rate as
\begin{equation}A=\frac{1}{2}\ln \frac{\widetilde{R}_{F}}{\widetilde{R}_{F}-r}+\frac{1}{2}\ln \frac{\widetilde{R}_{B}+r}{\widetilde{R}_{B}}=\frac{1}{2}\left(\frac{1}{\widetilde{R}_{F}}+\frac{1}{\widetilde{R}_{B}}\right)r+O(r^{3}),\label{Eq-homoauto-simplyAf}\end{equation}
and
\begin{equation}\sigma \equiv J_{m}A=\frac{1}{2}\left(\frac{1}{\widetilde{R}_{F}}+\frac{1}{\widetilde{R}_{B}}\right)r^{2}+O(r^{4}),\label{Eq-homoauto-simplysigma}\end{equation}
respectively. Surprisingly, from (\ref{Eq-homoauto-r}) and (\ref{Eq-homoauto-simplyRfRb}), we can prove ${1}/{\widetilde{R}_{F}}+{1}/{\widetilde{R}_{B}}=\sum_{i=1}^{N}({1}/{R_{F}^{i}}+{1}/{R_{B}^{i}})+O(r^{2})$. Thus (\ref{Eq-homoauto-simplyAf}) and (\ref{Eq-homoauto-simplysigma}) may be further transformed into
\begin{equation}A=\frac{1}{2}\sum_{i=1}^{N}\left(\frac{1}{R_F^i}+\frac{1}{R_B^i}\right)r+O(r^{3}),\end{equation}
and
\begin{equation}\sigma \equiv J_{m}A=\frac{1}{2}\sum_{i=1}^{N}\left(\frac{1}{R_F^i}+\frac{1}{R_B^i}\right)r^{2}+O(r^{4}),\end{equation}
respectively.

As an example, we will give the reducing procedure from the single-level quantum dot engine depicted in Sec.~\ref{Sec-homoauto-quantum}, which contains two elementary processes ($\mathcal{R}^{h}$ and $\mathcal{R}^{c}$), into a reduced model containing only one effective process ($\widetilde{\mathcal{R}}$).
As depicted in \cite{Esposito2009EPL}, when the engine operates in the steady state, the forward and backward flows in the process between the hot reservoir and the quantum dot as well as those in the process between the quantum dot and the cold reservoir could be expressed as
\begin{equation}R_{F}^{h}=\frac{a_{h}(1-f_{h})+a_{c}(1-f_{c})}{a_{h}+a_{c}}a_{h}f_{h},~~R_{B}^{h}=\frac{a_{h}f_{h}+a_{c}f_{c}}{a_{h}+a_{c}}a_{h}(1-f_{h});\label{Eq-homoauto-simplyRfRbh}\end{equation}
and
\begin{equation}R_{F}^{c}=\frac{a_{h}f_{h}+a_{c}f_{c}}{a_{h}+a_{c}}a_{c}(1-f_{c}),~~R_{B}^{c}=\frac{a_{h}(1-f_{h})+a_{c}(1-f_{c})}{a_{h}+a_{c}}a_{c}f_{c},\label{Eq-homoauto-simplyRfRbc}\end{equation}
respectively, with $f_{h}\equiv 1/[1+e^{(\varepsilon-\mu_{h})/T_{h}}]$ and $f_{c}\equiv 1/[1+e^{(\varepsilon-\mu_{c})/T_{c}}]$. $a_{h}$ (or $a_{c}$) is a coefficient in process $\mathcal{R}^{h}$ (or $\mathcal{R}^{c}$).

Because there are only two elementary processes in the single-level quantum dot model, the second and higher order terms in (\ref{Eq-homoauto-simplyRfRb}) is vanishing. Substituting (\ref{Eq-homoauto-simplyRfRbh}) and (\ref{Eq-homoauto-simplyRfRbc}) into (\ref{Eq-homoauto-simplyRfRb}), we derive the reciprocal of effective flows:
\begin{eqnarray}\frac{1}{\widetilde{R}_{F}}&=&\frac{a_{h}^{2}f_{h}(1-f_{h})+a_{c}^{2}f_{c}(1-f_{c})+2a_{h}a_{c}f_{h}(1-f_{c})-(a_{h}+a_{c})r}{a_{h}a_{c}(a_{h}f_{h}+a_{c}f_{c})[1-(a_{h}f_{h}+a_{c}f_{c})/(a_{h}+a_{c})]f_{h}(1-f_{c})},\nonumber\\
\frac{1}{\widetilde{R}_{B}}&=&\frac{a_{h}^{2}f_{h}(1-f_{h})+a_{c}^{2}f_{c}(1-f_{c})+2a_{h}a_{c}f_{c}(1-f_{h})+(a_{h}+a_{c})r}{a_{h}a_{c}(a_{h}f_{h}+a_{c}f_{c})[1-(a_{h}f_{h}+a_{c}f_{c})/(a_{h}+a_{c})]f_{c}(1-f_{h})}.\label{Eq-homoauto-simplyRfRbe}\end{eqnarray}
Then, with consideration of $\alpha=a_{h}a_{c}/(a_{h}+a_{c})$ and $r=\alpha (f_{h}-f_{c})$, we could verify $\widetilde{R}_{F}-\widetilde{R}_{B}=r$ and $\sigma=(\widetilde{R}_{F}-\widetilde{R}_{B})\ln \widetilde{R}_{F}/\widetilde{R}_{B}$ are exactly satisfied in this reduced model. Of course, both (\ref{Eq-homoauto-simplyAf}) and (\ref{Eq-homoauto-simplysigma}) still hold.

Further, with the consideration of (\ref{Eq-model-Xt}), (\ref{Eq-homoauto-quantumShSc}), (\ref{Eq-homoauto-quantumbeta}) and (\ref{Eq-homoauto-quantumXm}), $f_{h}\equiv 1/[1+e^{(\varepsilon-\mu_{h})/T_{h}}]$ and $f_{c}\equiv 1/[1+e^{(\varepsilon-\mu_{c})/T_{c}}]$ may be expressed as
\begin{equation}f_{h}=\frac{1}{1+e^{(\xi-s_{h}X_{m}/\beta)(\bar{\beta}-X_{t}/2)}},~\mathrm{and}~f_{c}=\frac{1}{1+e^{(\xi+s_{c}X_{m}/\beta)(\bar{\beta}+X_{t}/2)}},\label{Eq-homoauto-simplyfhfc}\end{equation}
respectively, with $\bar{\beta}=(1/T_{h}+1/T_{c})/2$ and $\xi=\varepsilon -\mu$. Substituting (\ref{Eq-homoauto-simplyfhfc}) into (\ref{Eq-homoauto-simplyRfRbe}), through tedious calculations we derive
\begin{equation}\hspace{-10mm}\frac{1}{\widetilde{R}_{F}}+\frac{1}{\widetilde{R}_{B}}=\frac{8}{\alpha}\cosh^{2}(\bar{\beta}\xi /2)\left[1-\frac{\tanh(\bar{\beta}\xi /2)}{2}(s_{h}-s_{c})X_{m}\right]+O(X_{m}^{2},X_{t}^{2}),\label{Eq-homoauto-simplyRfplusRb}\end{equation}
with consideration of $\bar{\beta}/\beta =1+O(X_{t})$ and $a_{h}a_{c}/(a_{h}+a_{c})=\alpha$. Then, combining (\ref{Eq-homoauto-simplyAf}) and (\ref{Eq-homoauto-simplyRfplusRb}), we can achieve the nonlinear constitutive relation (\ref{Eq-homoauto-quantumJm2}) for the single-level quantum dot heat engine again.

\section{Conclusion}
In this work, we first investigated the hidden symmetries existing in a broad class of heat engines. In cyclic heat engines, this hidden symmetry may be characterized as the duality of the heat exchanges between the engine and two reservoirs under the parity-time transformation [$\mathcal{PT}$-symmetry (\ref{Eq-homo-cyclicPT})]. Based on a generic model of nonequilibrium cyclic heat engine and this $\mathcal{PT}$-symmetry we derived generic nonlinear constitutive relation (\ref{Eq-homocyclic-Jm}) for cyclic heat engines. In autonomous heat engines, the hidden symmetry may be characterized as the duality of the forward and backward flows under parity inversion [$\mathcal{P}$-symmetry (\ref{Eq-homo-autoP})], which leads to formal expression (\ref{Eq-homo-autoRfRb}) of the forward and backward flows. By applying this formal expression in a generic multi-step autonomous heat engine, we derived nonlinear constitutive relation (\ref{Eq-homoauto-Jm}) for autonomous heat engines. The hidden symmetries as well as the nonlinear constitutive relations are all confirmed by typical heat engines in the literature that we have known. Besides, we also proved that the multi-step autonomous heat engine could be reduced to an effective autonomous engine containing only single transmission process from the perspective of entropy production.

\section{Acknowledgement}
The authors are grateful to financial support from the
National Natural Science Foundation of China (Grant No. 11322543).

\end{document}